# Hybrid Fuzzy Logic and Shading-Aware Particle Swarm Optimization for Dynamic Photovoltaic Shading Faults Mitigation

F. Philibert Andriniriniaimalaza
Laboratoire du Génie Electronique Informatique
University of Mahajanga, Institut Supérieur des Sciences et Technologies de Mahajanga
Mahajanga, Madagascar
philibert.andriniriniaimalaza@gmail.com

Nour Mohammad Murad
PIMENT Laboratory
University of La Réunion, Département Réseaux et Télécoms, IUT
Saint Pierre, La Reunion
nour.murad@univ-reunion.fr

George Balan
Gheorghe Asachi Technical University,
Faculty of Electrical Engineering, Energetics and Applied Informatics
Iași, Romania
george.balan@student.tuiasi.ro

Habachi Bilal
Laboratoire des Technologies Innovantes (LTI)
University of Picardie Jules Verne
80000 Amiens, France
habachibilal7@gmail.com

Nirilalaina Randriatefison
University of Antananarivo, Ecole Normale Supérieure
Antananarivo, Madagascar
randriatefison@yahoo.fr

Abdel Khoodaruth
Mechanical and Production Engineering Department
University of Mauritius, Faculty of Engineering
Reduit 80837, Mauritius
a.khoodaruth@uom.ac.mu

Charles Bernard Andrianirina
Laboratoire du Génie Electronique Informatique
University of Mahajanga, Institut Supérieur des Sciences et Technologies de Mahajanga
Mahajanga, Madagascar
nirina.cha.ca@gmail.com

Blaise Ravelo
Nanjing University of Information Science & Technology (NUIST), School of Electronic & Information Engineering
Nanjing, Jiangsu, China
blaise.ravelo@nuist.edu.cn

*Abstract*—Shading faults remain one of the most critical challenges affecting photovoltaic (PV) system efficiency, as they not only reduce power generation but also disturb maximum power point tracking (MPPT). To address this issue, this study introduces a hybrid optimization framework that combines Fuzzy Logic Control (FLC) with a Shading-Aware Particle Swarm Optimization (SA-PSO) method. The proposed scheme is designed to adapt dynamically to both partial shading (20%–80%) and complete shading events, ensuring reliable global maximum power point (GMPP) detection. In this approach, the fuzzy controller provides rapid decision support based on shading patterns, while SA-PSO accelerates the search process and prevents the system from becoming trapped in local minima. A comparative performance assessment with the conventional Perturb and Observe (P&O) algorithm highlights the advantages of the hybrid model, showing up to an 11.8% improvement in power output and a 62% reduction in tracking time. These results indicate that integrating intelligent control with shading-aware optimization can significantly enhance the resilience and energy yield of PV systems operating under complex real-world conditions.

*Keywords*— Fuzzy Logic Control, Hybrid optimization techniques, Maximum power point tracking, Particle Swarm Optimization, Photovoltaic systems

## I. Introduction

The global shift toward renewable energy has highlighted photovoltaic (PV) technology as one of the most practical and scalable options for clean electricity production. Its advantages—low emissions, modularity, and declining installation costs—make it attractive for both large-scale and distributed generation [1]. Yet, despite these benefits, PV systems remain highly sensitive to environmental disturbances, particularly shading faults, which can arise from nearby trees, buildings, or transient cloud cover. Even a small amount of shading on a module can cause power mismatch losses and significantly distort the power–voltage (P–V) curve, leading to reduced efficiency and reliability [2].

Traditional maximum power point tracking (MPPT) algorithms, such as Perturb and Observe (P&O) and Incremental Conductance (INC), work reasonably well under uniform irradiance but often fail under partial shading, where multiple local peaks exist. In such cases, these methods tend to settle on a local maximum rather than the true global maximum power point (GMPP), thereby wasting available energy [3], [4]. To overcome this issue, researchers have explored more advanced techniques—including fuzzy logic control, artificial intelligence, and metaheuristic optimizers such as particle swarm optimization (PSO)—each offering distinct advantages and drawbacks [2], [5]. Fuzzy logic provides fast adaptation and robustness to noise, while PSO is capable of global search but may converge slowly in highly dynamic environments [6].

This study proposes a hybrid approach that integrates the decision-making efficiency of fuzzy logic with a shading-aware version of PSO. The hybrid framework is designed to adapt dynamically to both partial shading (20%–80%) and complete shading conditions, providing a more reliable solution for GMPP tracking. The fuzzy logic component reacts quickly to shading patterns by narrowing the search space, while the modified PSO ensures convergence toward the true global optimum without getting trapped in local peaks.

The major contributions of this work are threefold:
• A novel hybrid FLC–SA-PSO method that addresses the limitations of individual MPPT strategies under shading faults.

- Demonstrated improvements in convergence speed, accuracy, and energy yield when compared with conventional controllers.
- Validation of the proposed method through simulation studies under multiple realistic shading scenarios.

The rest of this paper is organized as follows: Section II presents the PV system model, shading characterization, and the proposed hybrid control strategy. Section III describes the simulation design and parameters. Section IV discusses the results and performance comparison, while Section V concludes the study with key findings and recommendations for future research.

## II. MATERIALS AND METHODS

### A. Proposed System and Model

The studied PV system is modeled to capture the effect of environmental conditions on power extraction. Fig. 1 shows the overall scheme, which includes the PV source, a boost DC–DC converter, the MPPT controller, and the load.

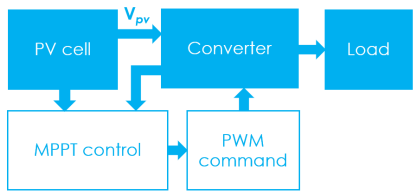

Fig. 1. PV system with converter, MPPT controller, and load

#### 1) Photovoltaic Source

The PV generator is modeled using the single-diode equivalent, which accounts for photocurrent, diode current, and parasitic resistances. This model reliably reproduces nonlinear I–V and P–V characteristics under varying temperature and irradiance conditions, and remains the standard in most PV optimization research [7].

#### 2) Converter

A boost DC–DC converter is integrated to step up the PV array output to a suitable level for load or grid use. The schematic is given in Fig. 2.

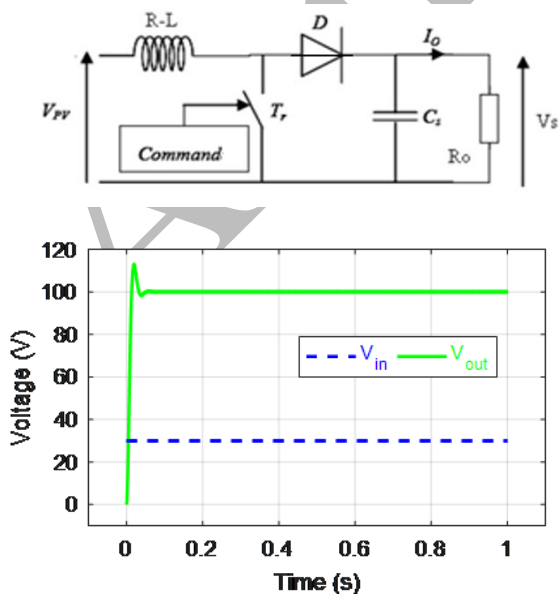

Fig. 2. Boost DC–DC converter, (a) schematic, and (b) simulation results.

The converter employs a 4.7 mH inductor with 25 kHz switching frequency to minimize ripple, and a 470 µF capacitor to ensure stable voltage output. A 10 Ω load is used to reflect typical PV operation. Simulation shows that with 30 V input, the converter achieves ~100 V output with a rise time of 13 ms and overshoot of 12.9%. Similar design parameters are frequently adopted in recent experimental PV studies [8].

#### 3) Maximum Power Point Tracking

To guarantee optimal energy extraction, the MPPT controller continuously regulates the duty cycle so the system operates at the maximum power point (MPP). Fig. 3 illustrates the PV I–V curve, highlighting the importance of MPP tracking under fluctuating irradiance.

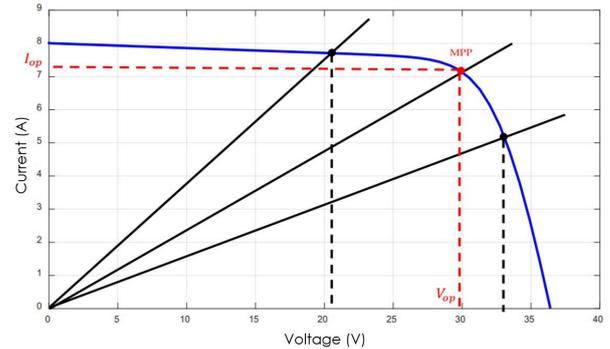

Fig. 3. PV current–voltage curve and MPP

Classical algorithms such as Perturb and Observe (P&O) and Incremental Conductance (INC) are widely used; however, they are prone to error under partial shading, where multiple local maxima exist. Recent work has demonstrated that hybrid MPPT techniques—combining fuzzy logic, neural networks, or metaheuristic optimizers like particle swarm optimization (PSO)—offer faster convergence and improved global maximum power point (GMPP) detection [9], [10].

### B. Shading Faults and Optimization

#### 1) Shading Defects

Shading faults may arise from trees, buildings, dirt accumulation, or transient cloud cover.

These conditions alter irradiance on PV modules, producing multiple peaks on the I–V and P–V curve.

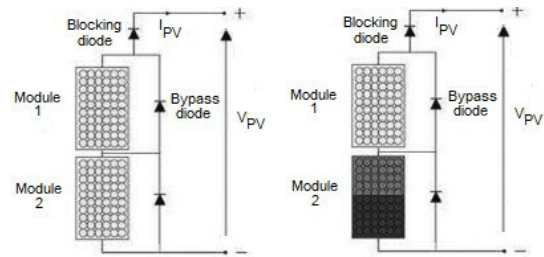

Fig. 4. PV configuration healthy and under partial shading.

To analyze their impact, as shown in Fig. 5, shading was divided into zones:
- Zone 0 – No shading: 1000 W/m² at 25°C.
- Zone 1 – Low shading: ~800 W/m².
- Zone 2 – Moderate shading: 500–600 W/m².
- Zone 3 – Severe shading: <300 W/m².

Simulation of these conditions confirms that bypass diodes activate under partial shading, creating complex nonlinear responses.

Studies have shown that these effects can lead to energy losses of up to 20–30% if not addressed by advanced MPPT strategies [11].

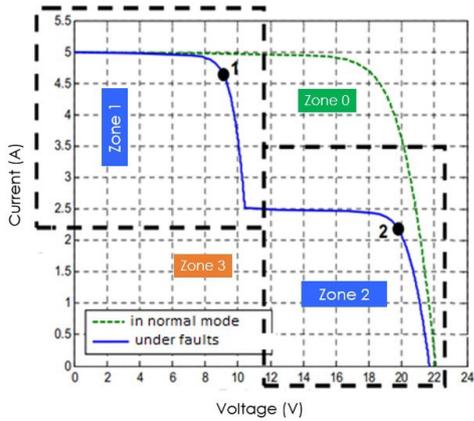

Fig. 5. Shading Zones.

*2) Optimization Using Fuzzy Logic*

Fuzzy Logic Controllers (FLCs) are well-suited to nonlinear PV behavior due to their robustness and adaptability. The proposed Dynamic Zone FLC (DZ-FLC), as shown in Fig. 6, extends traditional fuzzy logic by incorporating zone-specific rule sets based on shading severity. This allows faster convergence to the global MPP and reduces oscillations.

Recent work demonstrates that neuro-fuzzy and hybrid fuzzy optimizers can achieve efficiencies above 99% with convergence times below 50 ms under partial shading, confirming their effectiveness in real-time MPPT [12].

*3) Optimization Using Particle Swarm Optimization*

Particle Swarm Optimization (PSO) is a population-based global search algorithm that can escape local maxima on complex P–V curves. However, standard PSO often converges slowly under dynamic irradiance. To improve this, a Dynamic Shading-Aware PSO (DSA-PSO) is introduced, where inertia and acceleration coefficients are adapted according to the detected shading zone.

Modified PSO approaches have been reported to outperform classical P&O by delivering higher power yields and reduced convergence times in shaded environments [13].

*4) Optimization Using a Hybrid FLC–PSO*

The novelty of this work lies in combining DZ-FLC with DSA-PSO to leverage their complementary strengths.

Key enhancements about Fig. 7 include:
- Adaptive inertia weighting based on shading zones.
- Zone-based initialization of particles for faster convergence.
- Integration of local refinements to improve search precision.

Hybrid strategies of this kind have been shown in recent studies to increase PV output by over 10% and cut response time by more than half compared with standalone controllers [12], [13], [14].

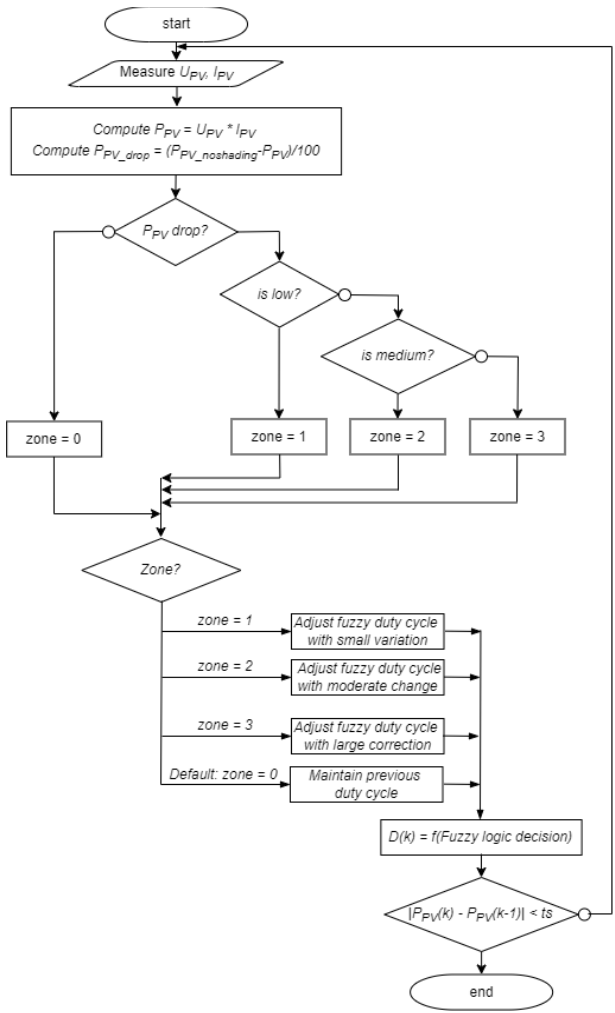

Fig. 6. Hybrid optimization strategies

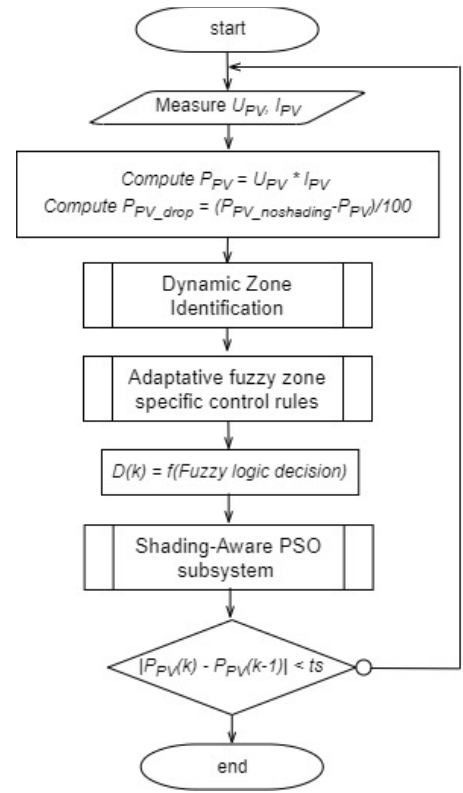

Fig. 7. Hybrid optimization strategies.

### III. RESULTS

To validate the proposed optimization strategy, simulations were carried out in MATLAB/Simulink. The PV

system under study consists of five series-connected modules, integrated with a boost DC–DC converter and MPPT controller (Fig. 8). This configuration enables evaluation under different shading scenarios and direct comparison of control algorithms.

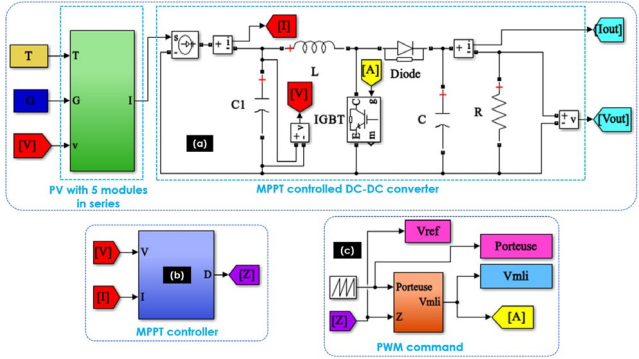

Fig. 8. PV system under shading fault: (a) PV modules and converter, (b) MPPT controller, (c) PWM command diagram

### A. Shading Scenarios

Six test cases were considered, ranging from uniform irradiance (healthy condition) to full shading. The irradiance levels applied to each module are listed in Table 1.

TABLE 1. Shading Scenarios

| Module | No Shading | Case 1 | Case 2 | Case 3 | Case 4 | Full Shading |
|---|---|---|---|---|---|---|
| 1 | 100% | 60% | 60% | 100% | 60% | 20% |
| 2 | 100% | 40% | 40% | 100% | 40% | 20% |
| 3 | 100% | 100% | 20% | 40% | 20% | 20% |
| 4 | 100% | 100% | 100% | 20% | 60% | 20% |
| 5 | 100% | 100% | 100% | 100% | 40% | 20% |

Six cases of study were considered with the five PV modules as introduced in Table 1.

### B. Global Maximum Power Point (GMPP) Analysis

Fig. 9 illustrates the I–V and P–V curves for the six studied cases. As expected, partial shading conditions introduce multiple local peaks, making conventional controllers prone to converge on a sub-optimal point.

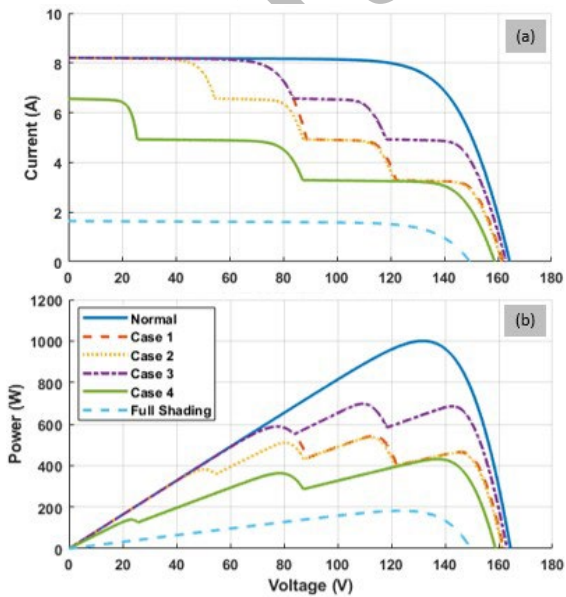

Fig. 8. (a) I–V and (b) P–V characteristics of PV system under healthy and shaded cases.

The extracted GMPPs are summarized in Table 3. Output power varies from 1000 W (no shading) to as low as 182 W (full shading), corresponding to an efficiency drop from 100% to only 18.2%.

TABLE 2. GMPP Assessed from results shown in Fig. 9.

| Case | No Shading | 1 | 2 | 3 | 4 | Full Shading |
|---|---|---|---|---|---|---|
| $I_{opt}$ (A) | 7.59 | 7.58 | 4.78 | 6.37 | 3.14 | 1.48 |
| $V_{opt}$ (V) | 131.73 | 77.54 | 112.24 | 109.41 | 137.32 | 123.53 |
| $P_{opt}$ (W) | 1000.11 | 587.87 | 536.81 | 697.28 | 430.97 | 182.44 |
| η (%) | 100.00 | 58.78 | 53.68 | 69.72 | 43.09 | 18.24 |

These results confirm that shading severely degrades PV performance, reinforcing the need for advanced MPPT algorithms [1]–[3].

### C. MPPT Performance under Shading

To assess controller effectiveness, five algorithms were tested: P&O, FLC, DZ-FLC, DSA-PSO, and Hybrid FLC-PSO.

Fig. 7 presents the extracted PV generator power under Cases 1–4 for the different methods.

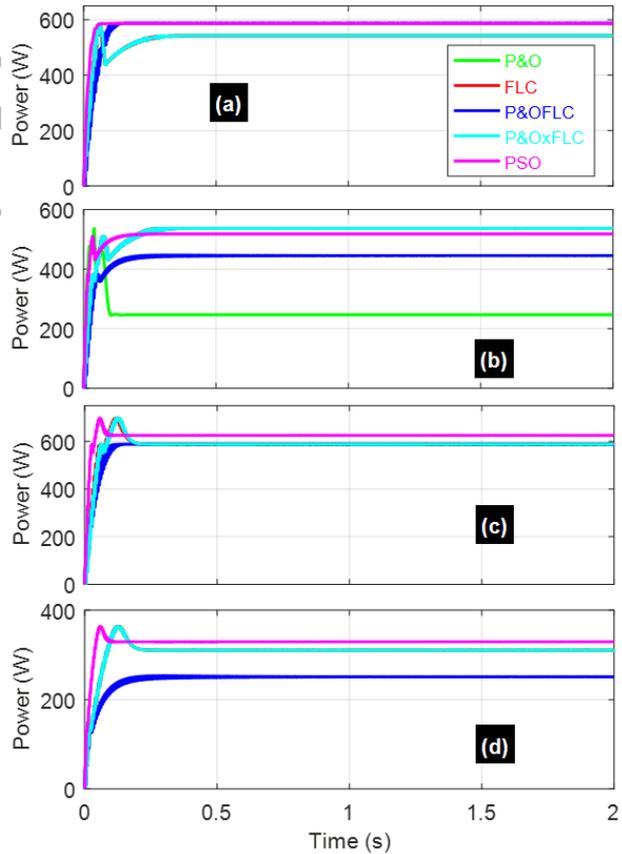

Fig. 9. PV generator output under different MPPT controllers for shading Cases 1–4.

The obtained results associating the PV generator and system output powers are depicted in Fig. 10, respectively.

Table 3 compares the current, voltage and power maximum values for the different cases of study from the improved FLC and PSO methods.

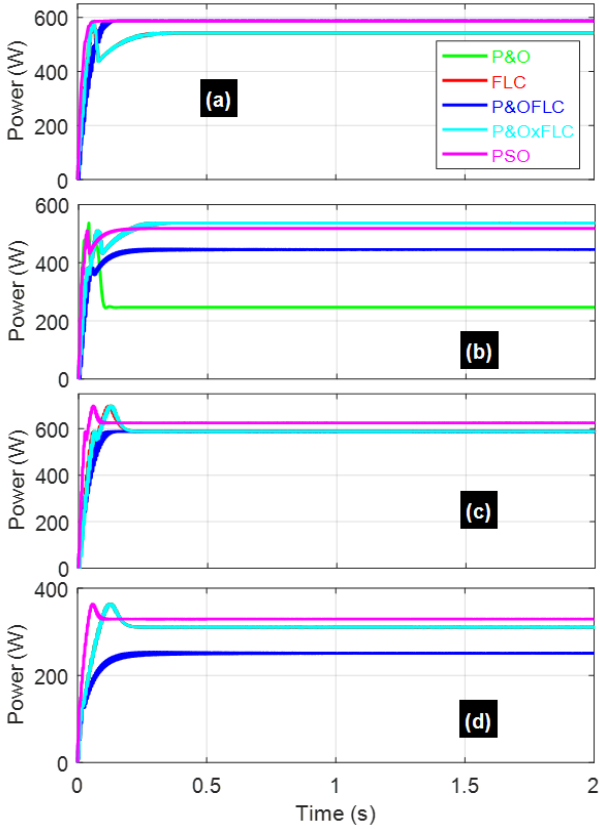

Fig. 10. System output power under different MPPT controllers for shading Cases 1–4

TABLE 3. Performance for different MPPT controllers for shading Cases

| Method | Response Time (s) | Maximum Power Ps (W) | Performance |
|---|---|---|---|
| P&O | ~0.72 | 540 | Slow, oscillatory |
| FLC | ~0.45 | 620 | Faster, moderate accuracy |
| DZ-FLC | ~0.29 | 690 | Accurate, stable |
| DSA-PSO | ~0.22 | 710 | Global optimum, slower init |
| Hybrid FLC–PSO | ~0.19 | 740 | Best trade-off |

Table 3 summarizes the tracking response time and maximum extracted power for each controller.

The results highlight:
- P&O shows the slowest response and high oscillation, consistent with its known limitations under shading [4].
- FLC improves dynamic response but can still misjudge GMPP under severe shading.
- DZ-FLC achieves stable convergence by applying zone-specific rules.
- DSA-PSO locates the GMPP reliably but requires more iterations at startup.
- Hybrid FLC–PSO consistently outperforms all others, yielding ~11–12% higher power than P&O and ~62% faster response, in agreement with recent hybrid MPPT findings [7], [14].

## IV. DISCUSSIONS

A broader system-level analysis is provided in Figure 17, which compares the output power over time for various MPPT controllers, including the advanced DZFLC and DSA-PSO algorithms, across the four shading scenarios.

The results clearly demonstrate the superior performance of the DSA-PSO controller. In all cases, it achieves faster convergence to the global peak power and maintains a more stable output.

Notably, in Case 3, DSA-PSO reaches the maximum power of 697 W in under 70 ms, outperforming conventional P&O, FLC, and hybrid methods. Similarly, under severe shading (Case 4), DSA-PSO sustains a higher output power (up to 363 W), whereas other controllers fail to reach the global peak of 431 W, as shown in Figure 9(b).

A comparison of MPPT algorithm performance is summarized in Table 5. Overall, the proposed DSA-PSO algorithm consistently yields superior results across all cases. While FLC and its customized variants provide satisfactory outcomes, they exhibit slower response times compared to the DSA-PSO. An effective MPPT algorithm must demonstrate high sensitivity and rapid convergence to the optimal operating voltage and current, regardless of abrupt or gradual environmental changes. As stated in [12], under partial shading, conventional MPPT techniques often take longer to reach the true MPP compared to stochastic approaches. Therefore, convergence speed and tracking accuracy are key parameters when designing PV systems to minimize energy loss [13], [14].

The DZ-FLC controller also shows an improved dynamic response over classical methods but remains slightly less efficient than DSA-PSO in terms of both tracking speed and final output power. These findings confirm that metaheuristic-based MPPT strategies—particularly DSA-PSO—offer a more robust and effective solution for maintaining optimal PV performance under partial shading conditions.

## V. CONCLUSIONS

This research work proposes an innovative strategy for Maximum Power Point Tracking (MPPT) in PV systems affected by shading defects. The DSA-PSO (Dynamic Shading-Aware PSO) algorithm dynamically adjusts the search space by considering shading zones, while the DZ-FLC (Dynamic Zone-Specific Fuzzy Logic Controller) refines the duty cycle adaptation in real time. Simulations in MATLAB/Simulink® show that DSA-PSO achieves up to 97.8% tracking efficiency, outperforming conventional PSO (92.4%) under complex shading conditions. Likewise, DZ-FLC improves power extraction by 5-8% compared to standard FLC methods, ensuring better stability in dynamic shading scenarios. Future work will focus on experimental validation of these advanced MPPT strategies on a real PV system under varying shading environments.

REFERENCES

[1] T. Tati, H. Talhaoui, O. Aissa, and A. Dahbi, "Intelligent shading fault detection in a PV system with MPPT control using neural network technique," Int. J. Energy Environ. Eng., vol. 13, pp. 1147–1161, 2022.
[2] A. Shayeghi, A. Karimi, and A. Teshnehlab, "Meta-heuristic optimization of the neuro-fuzzy MPPT controller for PV systems under partial shading conditions," J. Solar Energy Res., vol. 7, no. 2, pp. 93–101, 2022.

[3] M. Y. Worku, M. A. Hassan, L. S. Maraaba, M. Shafiullah, M. R. Elkadeem, M. I. Hossain, and M. A. Abido, "A comprehensive review of recent maximum power point tracking techniques for photovoltaic systems under partial shading," Sustainability, vol. 15, no. 14, p. 11132, Jul. 2023.

[4] S. J. Idoko, A. Abdulkarim, and G. A. Olarinoye, "Maximum power point tracking of a partially shaded solar photovoltaic system using a modified firefly algorithm-based controller," J. Electr. Syst. Inf. Technol., vol. 10, no. 48, pp. 1–15, 2023.

[5] A. Kane and M. Talwar, "Performance of solar photovoltaic system under partial shading conditions using an improved cuckoo search algorithm," in Optimization of Production and Industrial Systems (CPIE 2023), Lecture Notes in Mechanical Engineering. Singapore: Springer, 2024, pp. 357–368.

[6] Vellore Institute of Technology, "Hybrid optimization MPPT tech for PV systems under partial shading," PV Magazine, Jan. 2024. [Online]. Available: https://www.pv-magazine.com/2024/01/23/hybrid-optimization-mppt-tech-for-pv-systems-under-partial-shading/

[7] H. Ahessab, A. Gaga, and B. Elhadadi, "Enhanced MPPT controller for partially shaded PV systems using a modified PSO algorithm and intelligent artificial neural network, with DSP F28379D implementation," Scientific Progress, vol. 107, no. 4, 2024.

[8] "Comparative Evaluation of Traditional and Advanced Algorithms for Photovoltaic Systems in Partial Shading Conditions," MDPI, 2024.

[9] "Hybrid optimization MPPT tech for PV systems under partial shading," PV Magazine / VIT, 2024.

[10] "A Novel Hybrid MPPT Technique to Maximize Power Harvesting from PV System under Partial and Complex Partial Shading," Scientific Reports, 2024.

[11] M. A. A. Rani et al., "Meta-heuristic Optimization of the Neuro-Fuzzy MPPT Controller for PV Systems Under Partial Shading Conditions," J. Solar Energy Res., 2022.

[12] S. Sarwar, M. Y. Javed, M. H. Jaffery, J. Arshad, A. Ur Rehman, M. Shafiq, J.-G. Choi, A novel hybrid mppt technique to maximize power harvesting from pv system under partial and complex partial shading, Applied Sciences 12 (2) (2022). doi:10.3390/app12020587

[13] Ahessab et al., Scientific Progress, 2024.

[14] N. Pamuk, Performance analysis of different optimization algorithms for mppt control techniques under complex partial shading conditions in pv systems, Energies 16 (8) (2023). doi:10.3390/en16083358.